\documentclass[%
 reprint,
 amsmath,amssymb,
 aps,
floatfix,
]{revtex4-2}

\usepackage{graphicx}
\usepackage{dcolumn}
\usepackage{bm}

\newcommand{\beq}{\begin{equation}}
\newcommand{\eeq}{\end{equation}}

\newcommand{\eps}{\varepsilon}

\renewcommand{\rho}{\varrho}
\renewcommand{\theta}{\vartheta}
\renewcommand{\phi}{\varphi}

\renewcommand{\Im}{{\rm Im} \,}

\newcommand{\wegdamit}[1]{} 

\newlength{\lwveryfine}   
\newlength{\lwfine}   
\newlength{\lwnormal} 
\newlength{\lwthick}  
\newlength{\lwverythick}  

\begin{document}

\preprint{APS/123-QED}

\title{Threedimensional Alternating-Phase Focusing for Dielectric-Laser Electron Accelerators}

\author{Uwe Niedermayer}
 \email{niedermayer@temf.tu-darmstadt.de}
\author{Thilo Egenolf}
\author{Oliver Boine-Frankenheim}
\affiliation{%
Technische Universit\"at Darmstadt, Schlossgartenstrasse 8, D-64289 Darmstadt, Germany
}%

\date{\today}

\begin{abstract}
The concept of dielectric-laser acceleration (DLA) provides the highest gradients among breakdown-limited (nonplasma) particle accelerators and thus the potential of miniaturization. The implementation of a fully scalable electron accelerator on a microchip by twodimensional alternating phase focusing (APF), which relies on homogeneous laser fields and external magnetic focusing in the third direction, was recently proposed. 
In this Letter, we generalize the APF for DLA scheme to 3D, such that stable beam transport and acceleration is attained without any external equipment, while the structures can still be fabricated by entirely twodimensional lithographic techniques. 
In the new scheme, we obtain significantly higher accelerating gradients at given incident laser field by additionally exploiting the new horizontal edge. This enables ultra-low injection energies of about 2.5~keV ($\beta=0.1$) and bulky high voltage equipment as used in previous DLA experiments can be omitted. DLAs have applications in ultrafast time-resolved electron microscopy and -diffraction. Our findings are crucial for the miniaturization of the entire setup and pave the way towards integration of DLAs in optical fiber driven endoscopes, e.g., for medical purposes.
\end{abstract}


\maketitle
Dielectric Laser Acceleration (DLA) was already proposed in 1962~\cite{Lohmann1962ElectronWaves,Shimoda1962ProposalMaser}, however, first experiments came 50 years later~\cite{Breuer2013DielectricEffect, Peralta2013DemonstrationMicrostructure.} by means of femtosecond laser pulses and lithographic nanofabrication. 
Recent advances in ultrashort laser pulses have enabled demonstrations of damage-threshold and self phase-modulation limited record gradients approaching the GeV/m milestone for relativistic electrons~\cite{Wootton2016DemonstrationPulses, Cesar2018High-fieldAccelerator}.
At subrelativistic energies, driven by a table-top electrostatic pre-accelerator, gradients of 133~MeV/m~\cite{Yousefi2018SiliconFabrication} and 370 MeV/m~\cite{Leedle2015LaserStructure} were achieved in silicon pillar structures.
In order to create a functioning accelerator out of these impressive gradients, the interaction length needs to be increased while maintaining a stable 6D phase space confinement.
First approaches to beam dynamics in grating linacs were already made in the 1980'~\cite{Palmer1980ALinac, Kim1982SomeLinac, Pickup1985AFrequencies}. In 2012, Naranjo et al.~\cite{Naranjo2012StableHarmonics} showed that the nonsynchronous spatial (sub-) harmonics can provide a ponderomotive focusing force, which was later turned into an accelerator design for medium energy~\cite{Cesar2019AIIDLA}.
Another approach starts from showing that a periodic grating provides phase dependent forces which can be concentrated in one kick per grating cell~\cite{Niedermayer2017BeamScheme}. Per Panofsky-Wenzel theorem~\cite{Panofsky1956SomeFields}, this threedimensional kick is irrotational, and can thus be modeled as a (time-dependent) potential in the Hamiltonian. Removing the time-dependence is achieved by lattice integration of the linearized fields according to the Courant-Snyder (CS) theory~\cite{Courant1958TheorySynchrotron}.
An accelerator lattice design providing stable motion in the longitudinal and one transverse direction is obtained by an Alternating Phase Focusing (APF) arrangement of grating segments treated as thick lenses~\cite{Niedermayer2018Alternating-PhaseAcceleration}.
Full scalability of the APF-DLA is achieved by using a pulse-front-tilted (PFT) laser~\cite{Wei2017Dual-gratingLaser,Cesar2018OpticalAccelerator,Kozak2018UltrafastNanostructures} or an  on-chip photonic waveguide system~\cite{Hughes2018On-ChipAccelerators}, which in principle allows to work with arbitrary short pulses.

The equivalent magnetic focusing strength of an individual APF segment was predicted~\cite{Niedermayer2018Alternating-PhaseAcceleration} and experimentally demonstrated to be in the MT/m range~\cite{Black2019Laser-DrivenMicrostructures}.
Moreover, the energy modulation of a subrelativistic DLA can also be turned into ballistic bunching~\cite{Niedermayer2017DesigningChip,Black2019NetAccelerator,Schonenberger2019GenerationAcceleration}, however, the hereby created energy spread is too large to inject into a scalable APF-DLA accelerator. A proposed APF-based segmented buncher~\cite{Niedermayer2018Alternating-PhaseAcceleration} solves this problem and is currently being experimentally tested.
Another ongoing experiment is the demonstration of a periodically segmented APF transport channel~\cite{Niedermayer2019ChallengesAcceleration}, which is however limited in length by the Rayleigh range of the electron beam in the invariant direction. In~\cite{Niedermayer2018Alternating-PhaseAcceleration}, we proposed to overcome this limit by installing an external quadrupole magnet which constantly focuses the beam in the vertical direction.
A major challenge in the experimental realization of a fully scalable APF-DLA is the alignment of this external magnet and the sufficient homogeneity of the laser fields in the invariant vertical direction of the structure.

In this Letter, we generalize the APF-based confinement to both transverse directions. This enables to completely eliminate  external focusing devices in scalable DLAs of arbitrary length. Moreover, since the additional dimension provides an additional edge, the accelerating near-field is increased, allowing to push for lower injection energy at given aperture. Previously reported minimal injection energies of $9.6\,$keV~\cite{McNeur2016APhotoemitter}
required using higher harmonics (in~\cite{McNeur2016APhotoemitter} the 5th) and confinement was not attained. Here, we intend to stay at the first harmonic since it provides the slowest drop-off from the grating surface and thus the
highest center gradient. We show, that injection energies of 16.75~keV and 2.5~keV are attainable at laser wavelengths of 2~$\mu$m and 6~$\mu$m, respectively. Thus, bulky high voltage feedthroughs in the experimental chambers can be entirely omitted. 

\begin{figure}[t]
\centering
\includegraphics[width=0.5\textwidth]{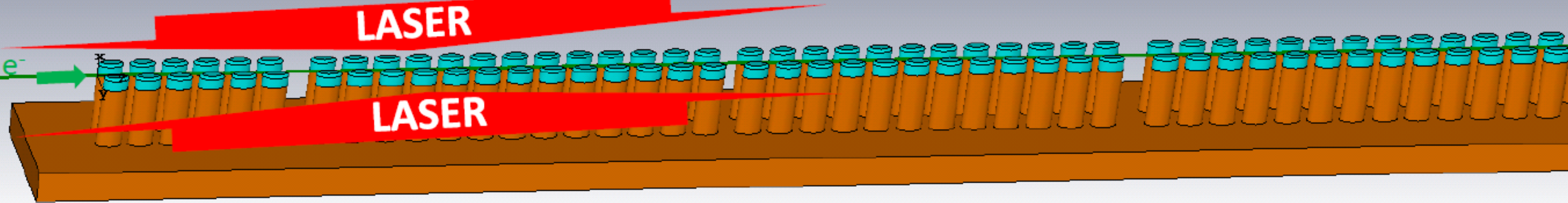}
\includegraphics[width=0.2\textwidth]{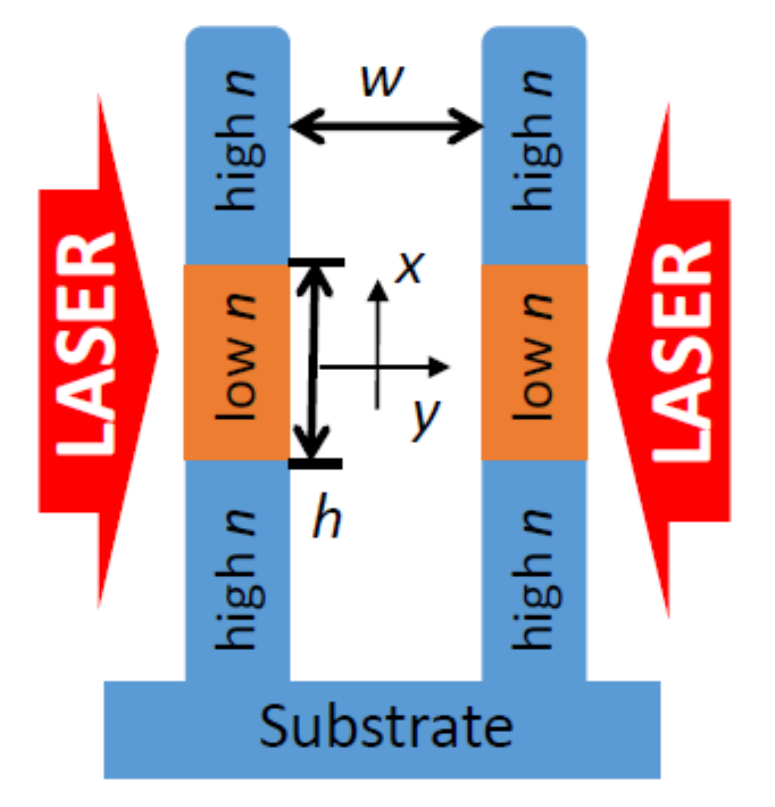}
\includegraphics[width=0.2\textwidth]{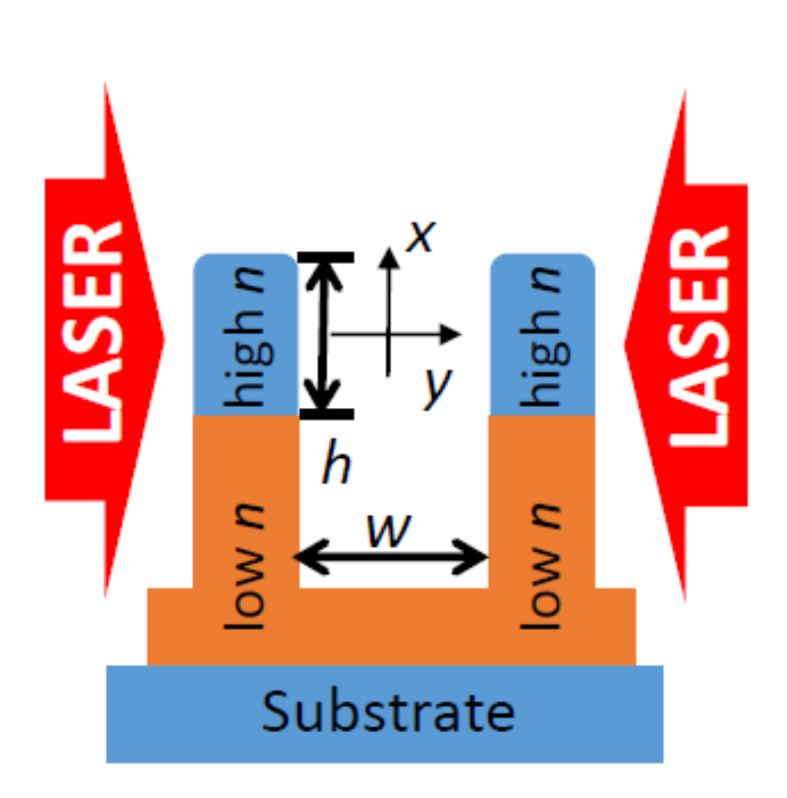}
\includegraphics[width=0.23\textwidth]{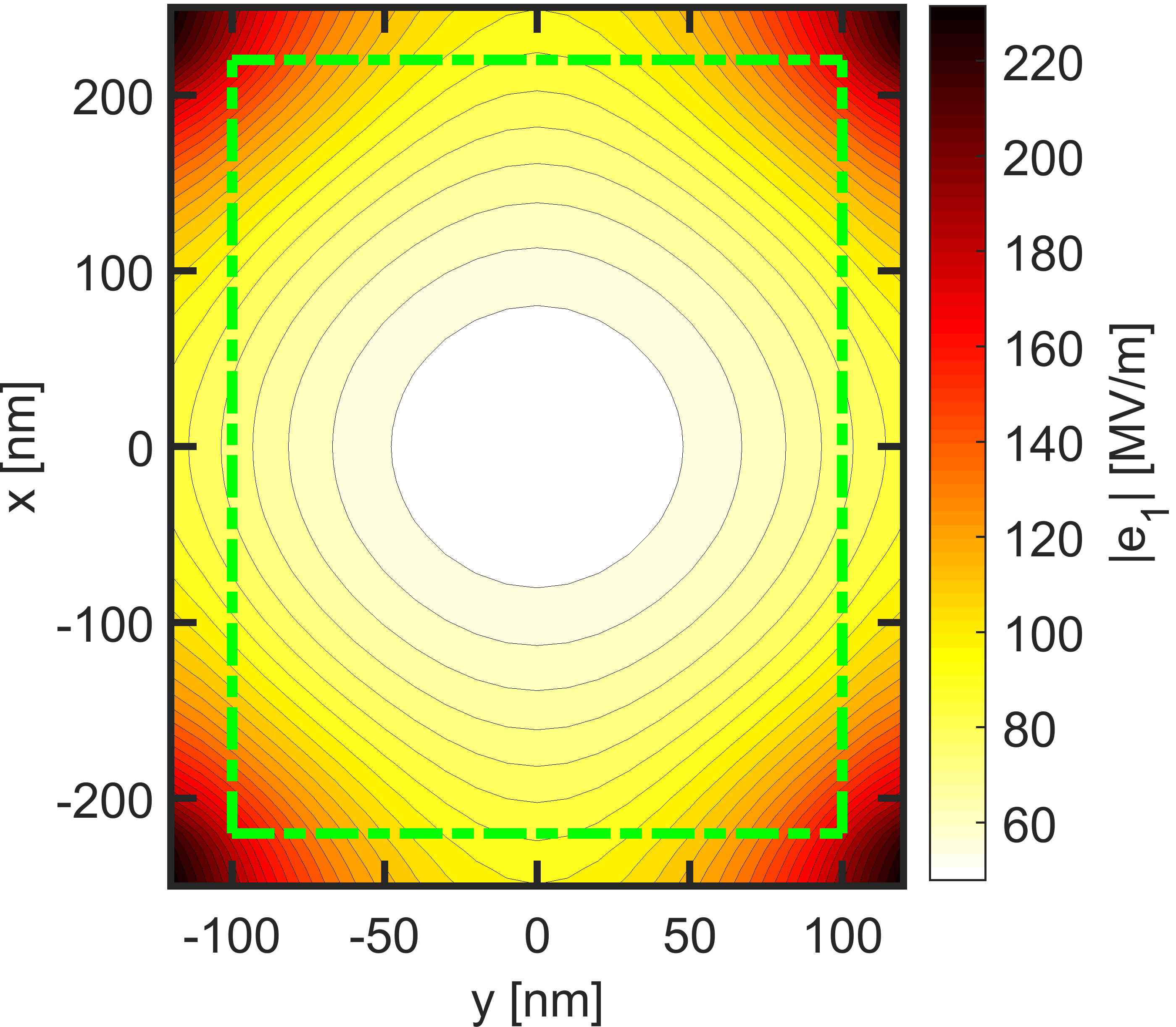}
\includegraphics[width=0.23\textwidth]{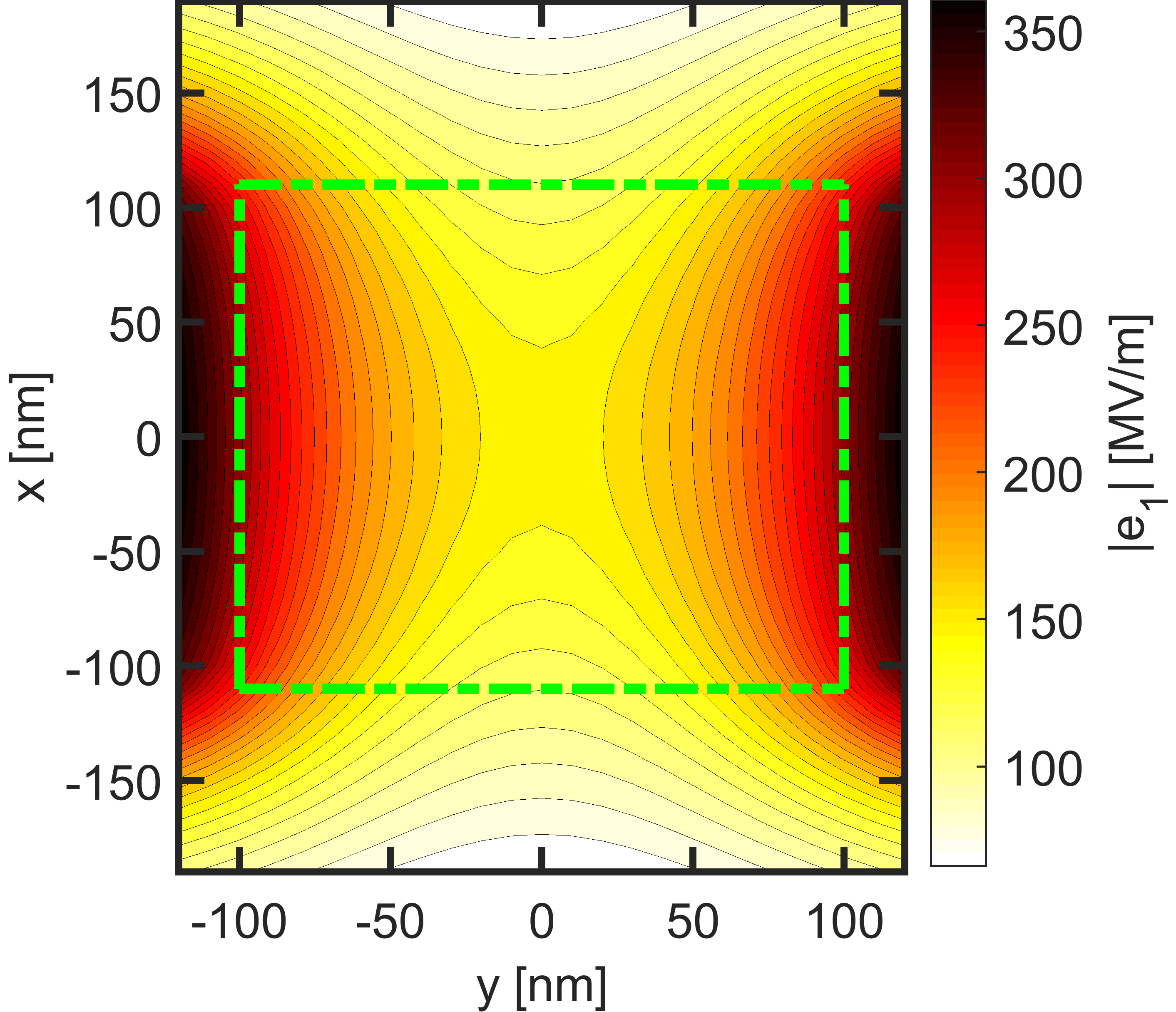}
\caption{Top: 3D APF-DLA based on SOI dual pillars.
Bottom: cross sections and $|e_1(x,y)|$ therein for
\textit{in-phase} APF (left) and \textit{counter-phase} APF (right, with approximation of $n_\mathrm{SiO_2}\approx 1$) and the beam channel $w\times h$ in green.}
\label{Fig:CrossSect}
\end{figure}

The 3D APF-based DLA can still be fabricated by 2D lithographic techniques.
The key idea is to work with two materials, exhibiting an as high as possible refractive index contrast, see Fig.~\ref{Fig:CrossSect}. Such technology is already commercially available in nano-electronics and -photonics, e.g. as Silicon-on-Insulator (SOI) wafers~\cite{2019Www.order.universitywafer.com} and has been used to demonstrate a waveguide driven DLA recently~\cite{Sapra2020On-chipAccelerator}.
The refractive indices at 2$\,\mu$m are~\cite{2019Www.refractiveindex.info} $n_\mathrm{Si}=3.67$ and $n_\mathrm{SiO_2}=1.44$, respectively.
At first, we make the approximation of $n_\mathrm{SiO_2}\approx 1$, which will be later refined. In other words, the oxide serves as just a building brick to construct 3D silicon structures by 2D lithography, where the layer thickness $h$ and the pillar semi-axis radii can attain single digit nanometer precisions. 

\begin{figure}[t]
\includegraphics[width=0.4\textwidth]{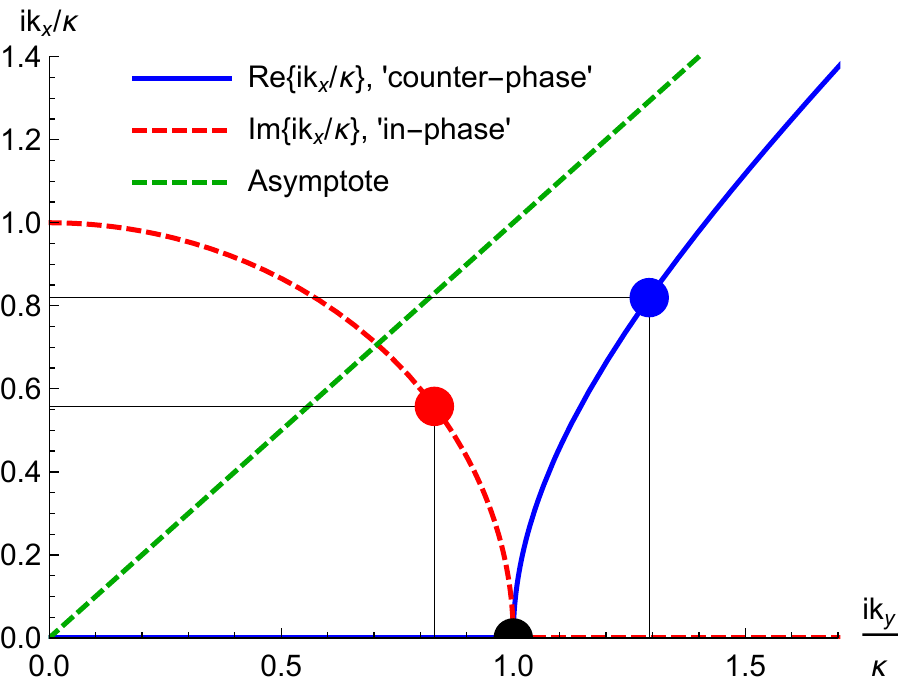}
\caption{Relation of $ik_x$ and $ik_y$. The black dot represents the two-dimensional APF scheme introduced in~\cite{Niedermayer2018Alternating-PhaseAcceleration}. The red and blue dots are examples of the \textit{in-phase} and \textit{counter-phase} APF scheme, respectively.
}
\label{Fig:Dispersion}
\end{figure}

\begin{figure}[b]
\includegraphics[width=0.4\textwidth]{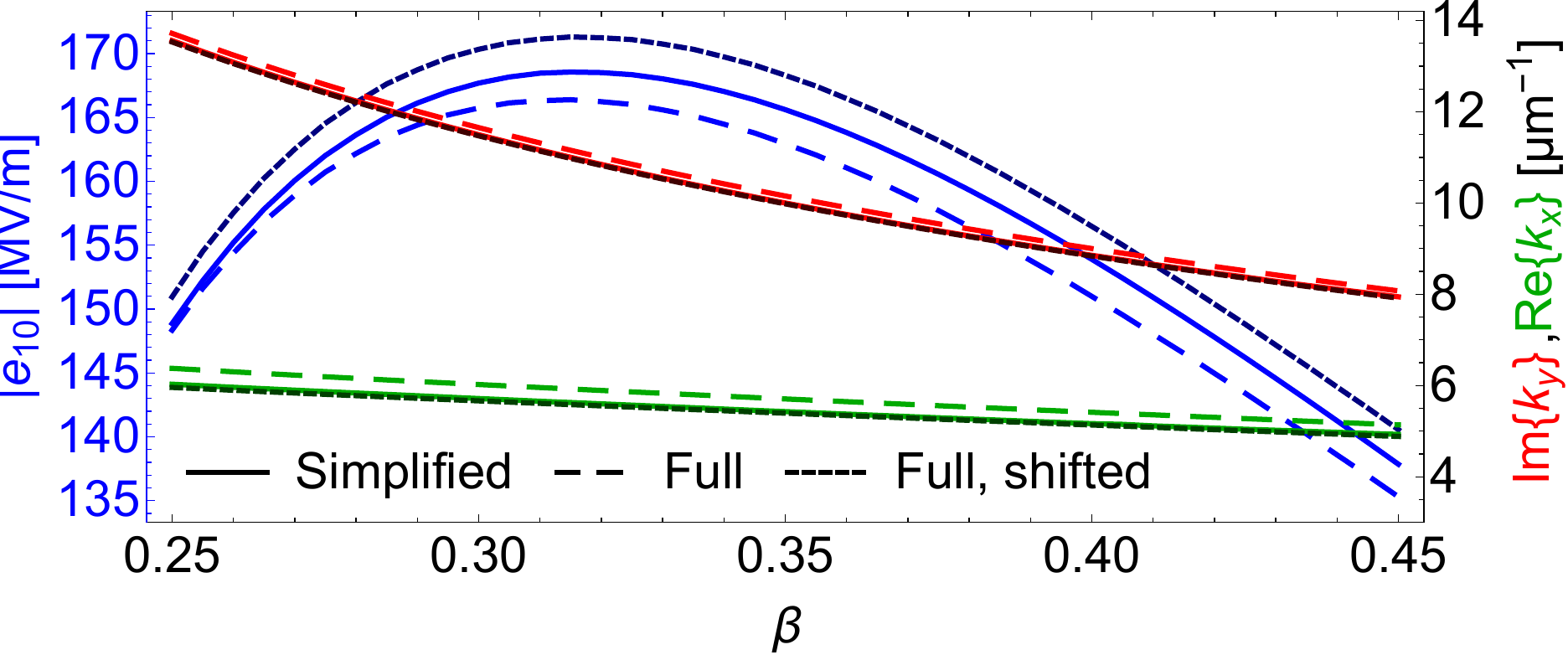}
\caption{Comparison of acceleration and focusing strength for simplified pillars (free floating), full pillars, and full pillars shifted by $\Delta x=10$\,nm towards the substrate for \textit{counter-phase} structures.}
\label{Fig:ComparisonFullSimplified}
\end{figure}

To model the electromagnetic fields in the (quasi-) periodic structures, we first look at the Helmholtz wave equation, in temporal frequency domain and Fourier series expanded in the longitudinal direction (see~\cite{Niedermayer2017BeamScheme}),
\begin{equation}
    \left[\triangle_\perp- k_z^2 +\omega^2/c^2 \right]e_1 (x,y)=0,
    \label{Eq:Helmholtz}
\end{equation}
where $\triangle_\perp$ is the transverse Laplacian, $k_z=\omega/(\beta c)$ is the longitudinal wave number of the synchronous mode $e_1$ to an electron traveling at speed $\beta c$ and $\omega=2\pi c/\lambda$. 
Equation~\ref{Eq:Helmholtz} is valid only in the vacuum of the beam channel. In contrast to conventional metallic accelerators, this creates the problem that boundary conditions, necessary to solve Eq.~\ref{Eq:Helmholtz}, are not available. We can however determine the dispersion relation from Eq.~\ref{Eq:Helmholtz} by $\partial_x\rightarrow -ik_x$, $\partial_y\rightarrow -ik_y$, and $\gamma^2=1/(1-\beta^2)$ as
\begin{equation}
    k_x^2+k_y^2= \frac{\omega^2}{c^2}-k_z^2= -\frac{\omega^2}{\beta^2\gamma^2 c^2}=:-\kappa^2,
    \label{Eq:Dispersion}
\end{equation}
which is plotted in Fig.~\ref{Fig:Dispersion}.
Instead of solving Eq.~\ref{Eq:Helmholtz}, we only need to determine $e_{10}=e_1(0,0)$ and one transverse wave number, which can be done numerically (we use CST~\cite{CST2019CSTSuite} as described in~\cite{Niedermayer2017BeamScheme}) for each individual grating cell with periodic boundary conditions in $z$-direction.
The transverse dependence of $e_1$ can then be written analytically as 
\begin{equation}
    e_1(x,y)=e_{10}\cosh(ik_x x) \cosh(ik_y y),
    \label{Eq:e1}
\end{equation}
which is numerically confirmed within 5\% in the channel $w\times h$ and plotted in Fig.~\ref{Fig:CrossSect} over a slightly larger range.
The assumption that the oxide can be neglected is crosschecked in Fig.~\ref{Fig:ComparisonFullSimplified}, where $|e_{10}|$, $k_x$, and $k_y$ are compared for a free floating simplified pillar, a full pillar with the origin centered in $h$, and a full pillar with the origin (i.e. the expansion point) shifted by 10 nm towards the substrate. Since the curves agree to sufficient accuracy, we will continue with the simplified pillars for brevity.

For the laser traveling in $\pm y$-direction and polarized in $z$-direction, $ik_y$ is always a purely real number. However, $ik_x$ can either be purely real or purely imaginary, see Fig.~\ref{Fig:Dispersion}. We will refer to these cases as \textit{in-phase} and \textit{counter-phase} scheme,
indicating whether or not both transverse planes are simultaneously focused. 

\begin{figure}[t]
\includegraphics[width=0.4\textwidth]{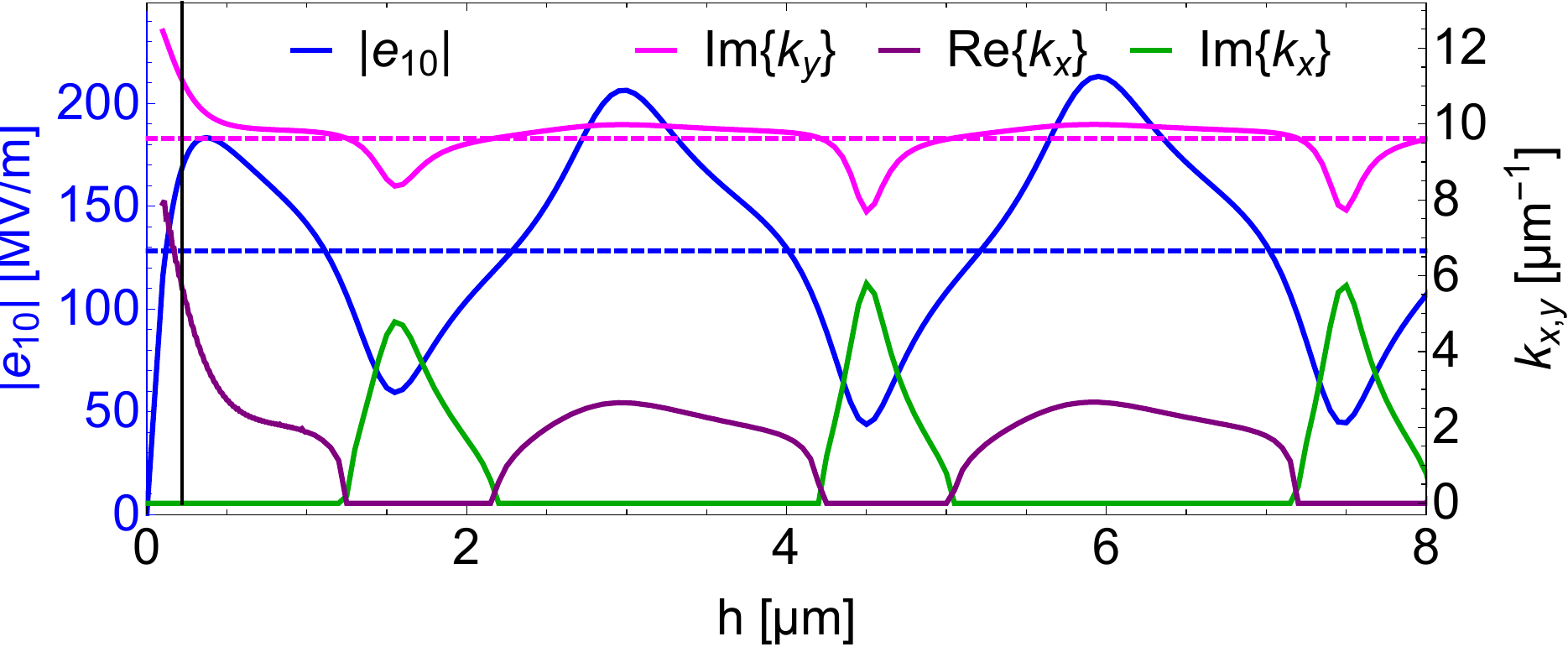}
\caption{Height parameter scan for a free-floating dual pillar (\textit{counter phase}) setup at $\beta=0.31$. Periodically reoccurring vertical eigenmodes make the twodimensional ($k_x=0$) case an exceptional case. The twodimensional case, obtained by using periodic boundaries in $x$, is plotted as dashed lines.}
\label{Fig:PillarHeight}
\end{figure}

With no loss of generality, we continue with the \textit{counter-phase} scheme only, since the structures are straightforward to fabricate on SOI wafers. 
The dependence of the wavenumbers $k_{x/y}$ as well as $e_{10}$ on the height $h$ of a free floating pillar
is plotted in Fig.~\ref{Fig:PillarHeight}. The dashed lines correspond to the 2D ($k_x=0$) case as introduced in~\cite{Niedermayer2018Alternating-PhaseAcceleration} (black dot in Fig.~\ref{Fig:Dispersion}).
The oscillation of $|e_{10}|$ is due to eigenmodes arising in the $x$-direction. Subsequently, the 2D case holds only for discrete values of $h$. These roots (and the corresponding maxima) of $k_x$ are not robust with respect to perturbations such as attaching the pillar to the oxide or to the substrate. Robust \textit{counter-phase} focusing behavior is however found below the first root of $k_x$, where also a 40\% enhancement of $|e_{10}|$ due to the finite height appears.
A reasonable thickness of $h=0.22\,\mu$m (commercially available~\cite{2019Www.order.universitywafer.com}) is chosen, slightly below this first maximum. A numerical parameter scan over $\beta$ using CST~\cite{CST2019CSTSuite} provides the results as depicted in Fig.~\ref{Fig:ComparisonFullSimplified}, where the entire curve is produced by an identical pillar design and only the cell length $\lambda_g$ is swept in order to always fulfill the Wideroe condition $\lambda_g=\beta\lambda$. 
Keeping the same pillar dimensions over a range of $\lambda_g$ requires compensating the phase drift, which is achieved by shifting the pillar center off the cell center by $\Delta z^{(n)}=\lambda_g^{(n)}(\arg(e_{10}^{(n)})-\arg(e_{10}^{(1)}))/2\pi$~\cite{Niedermayer2020SeeExamples.}.
This allows to use only about 4 or 5 different pillar designs to cover the entire range from extremely low energy to relativistic ($\beta\approx 1$) beams, which consists of in the order of $10^5$ pillar pairs.

With the above knowledge of the electromagnetic field, we proceed to the Hamiltonian $H=\Delta\vec P^2 /(2m_e\gamma)+V$, where $\Delta\vec P = (p_x,p_y,\Delta p_z/\gamma)^\mathrm{T}$ is the momentum deviation from the reference particle and $m_e$ is the electron rest mass.
The time-dependent potential $V$ reads generally (same procedure as in~\cite{Niedermayer2018Alternating-PhaseAcceleration}) as
\begin{equation}
V(x,y,s)=q\Im \{ k_z^{-1} e_1(x,y)e^{ik_z s}- is e_{10}e^{i\phi_s}\},
\label{Eq:Pot}
\end{equation}
where $q$ is the (negative) electron charge, $s$ is the relative longitudinal coordinate w.r.t. the laser phase,
and $\phi_s$ is the synchronous phase at which the reference particle gains energy according to the design acceleration ramp.
Tracking with the nonlinear kicks according to Eq.~\ref{Eq:Pot} can be performed using DLAtrack6D~\cite{Niedermayer2017BeamScheme}.
Expanding Eq.~\ref{Eq:Pot} to second order, Hamilton's equations provide the Hill's equations 
\begin{subequations}
\begin{align}
\Delta s'' &+ K_s \Delta s=0\\
y'' &+  K_y y=0\\
x'' &+  K_x x=0,
 \label{HillEqs}
\end{align}
\label{EqOfMotion}%
\end{subequations}
where $\Delta s= s- \lambda_g\phi_s/2\pi$. Due to the absence of first order terms in $V$, the linearized motion is decoupled.
\begin{figure}[b]
\centering
\includegraphics[width=0.45\textwidth]{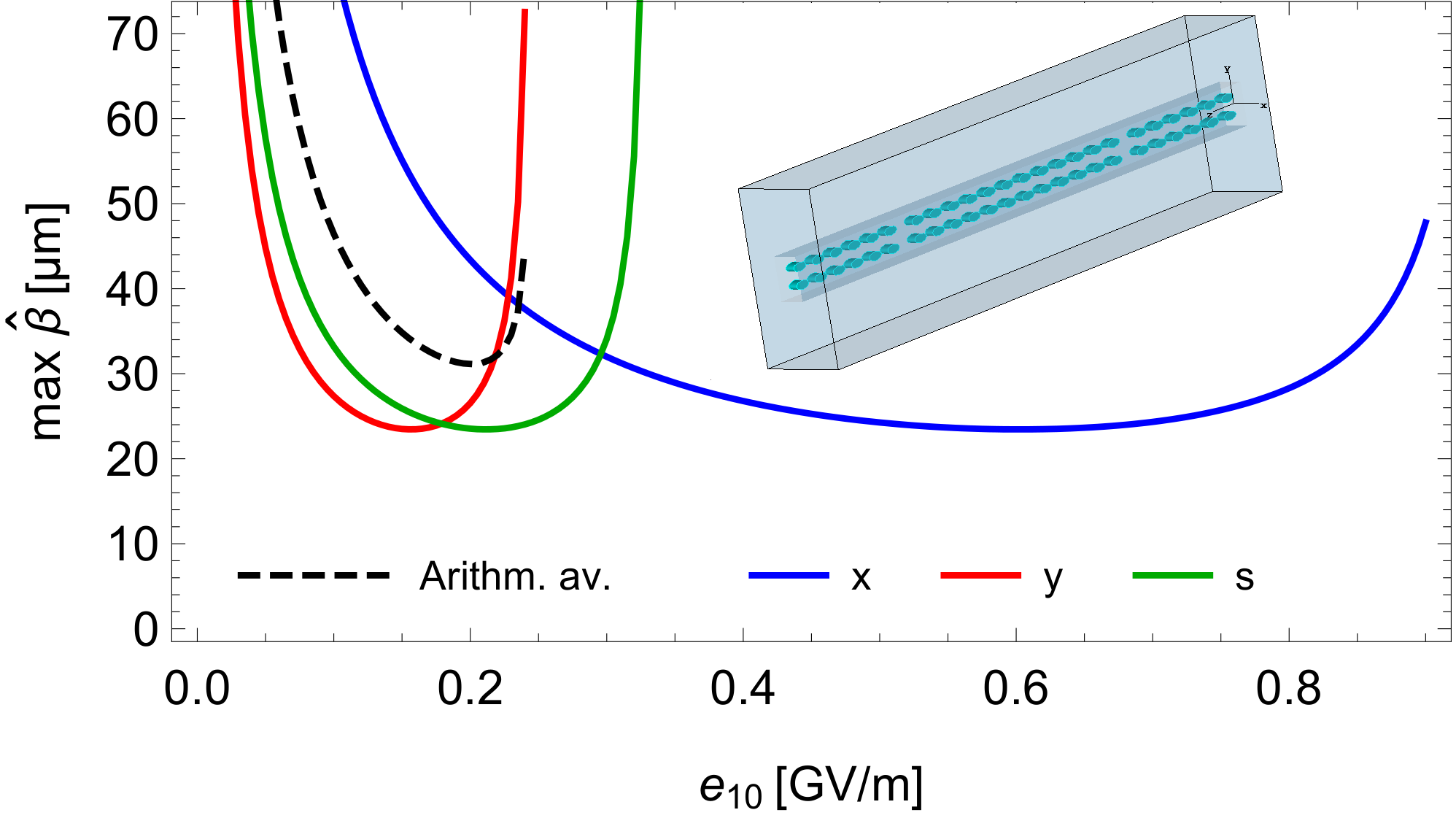}
\caption{Maximum of each $\hat\beta$-function computed from the eigenvalue problem. The inset shows the 3D model of the corresponding \textit{counter-phase} APF period.}
\label{Fig:Transport}
\end{figure}
\begin{figure*}[t]
\includegraphics[width=0.95\textwidth]{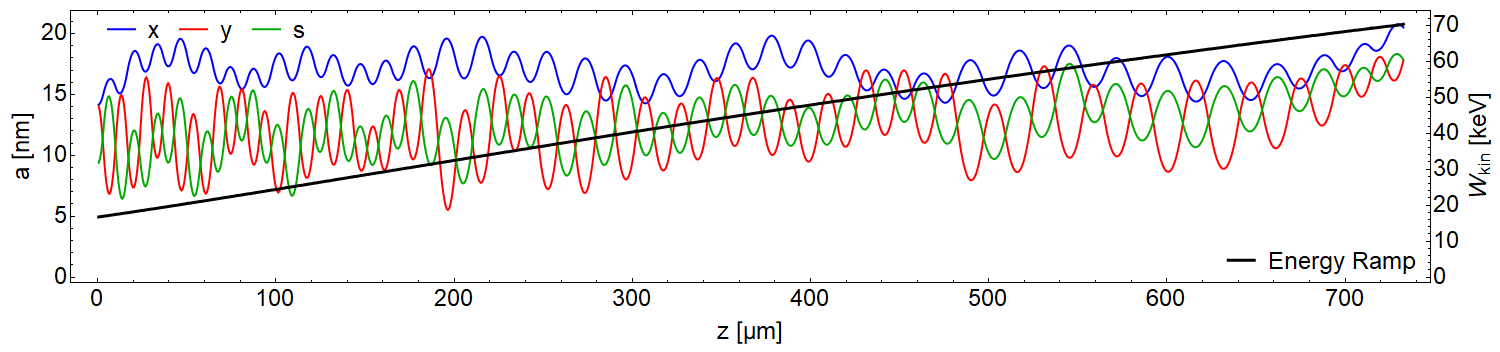}
\includegraphics[width=\textwidth]{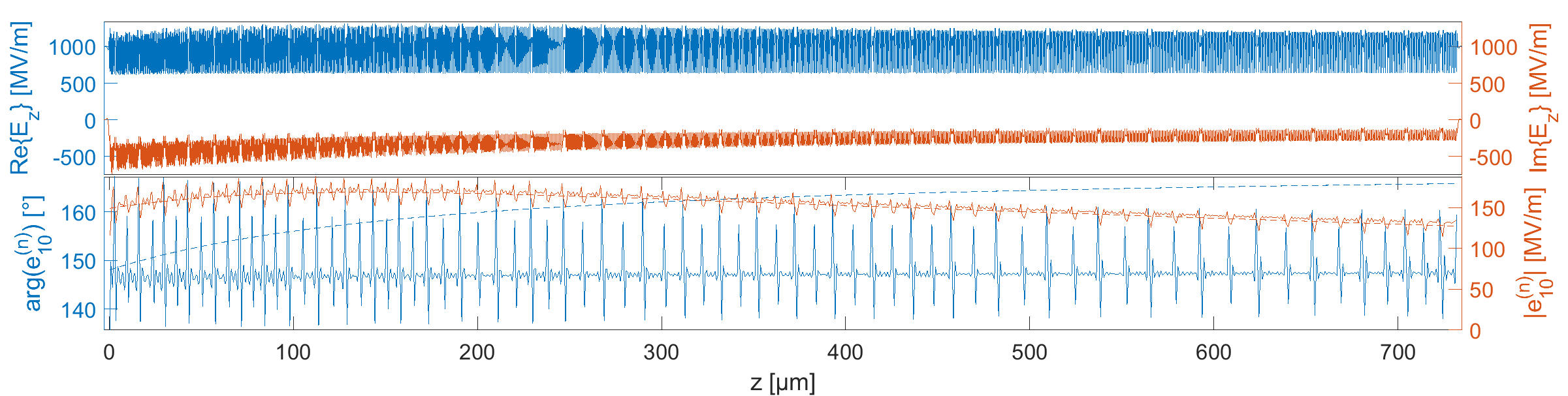}
\caption{Upper panel (design): envelopes for $\eps_n=2.5\,$pm and kinetic energy ramp.
Lower Panel (3D analysis): complex electric field $E_z(0,0,z)$ and spatial Fourier coefficients $e_{10}^{(n)}$ calculated from the full field windowed in each DLA cell $n$. The dashed lines represent the individually computed values of $e_{10}$ under periodic boundary conditions, which were used for the design.}
\label{Fig:Envelope}
\end{figure*} 
The focusing functions are
\begin{subequations}
\begin{align}
K_s&=-\frac{k_z^2}{\gamma^2}
\frac{ |qe_{10}|}{m_e\beta\gamma c \omega}\sin(\phi_s)\\
K_y&=(ik_y)^2\frac{ |qe_{10}|}{m_e\beta\gamma c \omega}\sin(\phi_s)\\
K_x&=(ik_x)^2\frac{ |qe_{10}|}{m_e\beta\gamma c \omega}\sin(\phi_s)
\end{align}
\label{HillK}%
\end{subequations}
and fulfill $K_x+K_y+K_s=0$ according to Eq.~\ref{Eq:Dispersion}, which reflects Earnshaw's theorem~\cite{Earnshaw1842OnEther}. Note that $K_s$ is the same as $K$ in the 2D scheme~\cite{Niedermayer2018Alternating-PhaseAcceleration} and the $s$ and $y$ planes are alternatingly focused by switching $\phi_s$ using fractional period drifts. The numerical value of $K_y$ is however different from the 2D case and thus all three frequencies (and subsequently phase advances) are disparate.

This general description of the motion is now turned into a functioning accelerator that provides 3D particle confinement by individual CS lattice integration~\cite{Courant1958TheorySynchrotron} in each plane $x,y,s$.
As in~\cite{Niedermayer2018Alternating-PhaseAcceleration}, the lattice functions in Eqs.~\ref{HillK} are converted to CS-functions (also called Twiss parameters) $\eta=(\hat\beta,\hat\alpha,\hat\gamma)^\mathrm{T}$ by solving the Twiss map eigenvector problem  $\eta_0=\mathbf{T}\eta_0$ for the initial values and subsequently mapping them to any other position. 
An example of a pure transport structure, which is strictly periodic, is shown in Fig.~\ref{Fig:Transport}. The parameter choice can be understood as a multi-objective optimization to find the minimum for the maxima of the $\hat\beta$ functions in all three dimensions, with a common APF cell length and a common incident laser field strength. A suitable choice on the resulting Pareto-front is the minimum of the arithmetic average of the three maxima of the $\hat\beta$ functions.

These individual cells are now combined to a full accelerator on a chip.
Initially we pick the injection energy and the synchronous phase, which is a compromise between desired acceleration gradient and required longitudinal focusing strength. The laser field strength is picked as slightly below the damage threshold fluence for a (curved-tilted) 100~fs pulse~\cite{Niedermayer2020SeeExamples.}.
A laser amplitude of 500 MV/m from each side and a synchronous phase of $\pm 60^{\circ} $ off-crest are chosen, leading to an average gradient of $G=\cos(\phi_s) \cdot|e_{10}^\mathrm{(norm)}|\cdot$500~MeV/m $\approx$73~MeV/m, where $e_{10}^\mathrm{(norm)}$ is the normalized synchronous mode coefficient. The acceleration ramp is determined by adding up the exact energy gains $|e_{10}^{(n)}|\lambda_{g}^{(n)}\cos(\phi_s)$ for each DLA cell $n$.

For a continuum of velocities $\beta$, the maps $\mathbf{T}_\mathrm{P}^{x,y,s}$ of each segment in each plane $x,y,s$ of the  
lattice are determined successively by solving eigenvalue problems for assumed periodic segments.
The length of each cell is taken as the arithmetic average of the minima of the maxima as indicated in Fig.~\ref{Fig:Transport}. 
Usually, a lattice obtained by simple matrix mapping of $\eta_0$ will exhibit growing $\hat\beta$-functions, due to cummulation of the small mismatch between two APF cells. Smooth and slowly growing $\hat\beta$-functions, such that the envelopes $a=(\hat\beta \eps)^{1/2}$ are non-growing, are obtained by manual correction of the segment lengths. Note that a slight growth of $\hat\beta$ is tolerable, since the emittance decreases by adiabatic damping according to $\eps=\eps_n/(\beta\gamma)$, where the normalized emittance $\eps_n$ is an invariant of the linearized motion. 

\begin{figure}[b]
\includegraphics[width=0.4\textwidth]{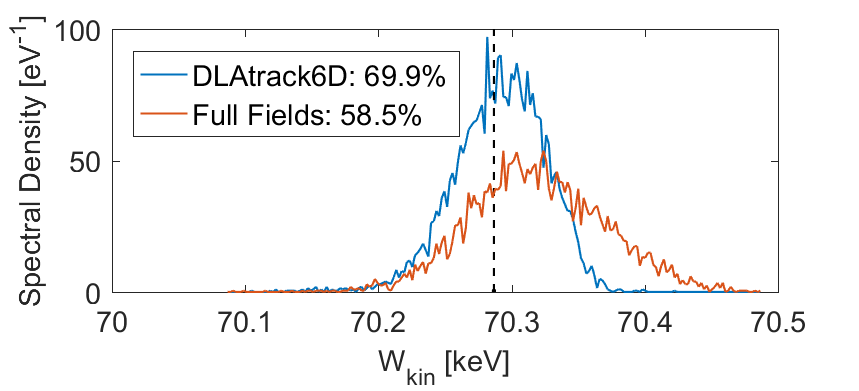}
\caption{Comparison of the final energy spectrum and throughput from a DLAtrack6D~\cite{Niedermayer2017BeamScheme} simulation vs. tracking simulation in the full laser fields using CST~\cite{CST2019CSTSuite}. The dashed vertical line is the design top energy.}
\label{Fig:ESpec}
\end{figure}

The resulting envelopes for a design for 16.75~keV ($\beta=0.25$) to 70~keV ($\beta=0.48$) are plotted in Fig.~\ref{Fig:Envelope}. After a full 3D field simulation~\cite{CST2019CSTSuite}, the complex field result along the channel is plotted below. Windowing this global field for each DLA cell allows a comparison between the $e_{10}^{(n)}$ in the entire accelerator to the individual $e_{10}^{(n)}$ initially computed under periodic boundary conditions (cf. Fig.~\ref{Fig:ComparisonFullSimplified}). As visible, the phase drift compensation keeps $\arg(e_{10}^{(n)})$ constant, but at the $\phi_s$-jumps a Gibbs phenonmenon is visible. Eventually this will have a detrimental effect on emittance preservation. The injection parameters were chosen as Gaussian distributions with geometric emittances $\eps_x=12$~pm, $\eps_x=7$~pm, and bunch length $\sigma_s=$5~nm with matched energy spread. These values are at the clipping point, where strong beam losses start to occur. The throughput and energy spread is shown in Fig.~\ref{Fig:ESpec} for one-kick-per-cell vs. 3D full field tracking. Similar results for 2.5~keV ($\beta=0.1$) to 16.75~keV ($\beta=0.25$) using $\lambda=6\,\mu$m are discussed in the Supplemental Material~\cite{Niedermayer2020SeeExamples.} and in principle, one could even start at a few eV
only, by using a Terahertz driver. 
However, lowering the injection energy poses a challenge to the robustness. Structure bandwidth, fabrication tolerances, and injection energy mismatch have to be controlled more precise. Normalized emittances in the single digit picometer range are available~\cite{Ehberger2015HighlyTip,Feist2017UltrafastBeam}, however, after the electrostatic pre-accelerator mostly higher values are reported (e.g.~\cite{Tafel2019FemtosecondTips}). This is due to nonlinear aberrations in the electrostatic lensing system. Our findings ease this problem significantly, since aberrations scale with the overall size of the system, which can be significantly reduced at lower injection energy. 

As confirmed by full 3D simulation, the 3D-APF-DLA scheme on SOI wafers is ready for experimental testing. Beyond the double sided lateral laser illumination one might also consider single beam top illumination or even pinched pure silicon pillars~\cite{Miao2020SurfaceThreshold}. More detailed theoretical studies are required to assess the effects of nonlinear and coupled particle-amplitude-dependent tune spreads, e.g., with the extended DLAtrack6D~\cite{Egenolf2019TrackingStructures}.

U.N. would like to thank Peyman Yousefi, Payton Broaddus, and Olav Solgaard for the discussions on SOI wafer processing.
This work is funded by the Gordon and Betty Moore Foundation under Grant No. GBMF4744 (ACHIP).

\bibliography{./references}


\end{document}



\title{Supplemental Material}

\author{Uwe Niedermayer}
 \email{niedermayer@temf.tu-darmstadt.de}
\author{Thilo Egenolf}
\author{Oliver Boine-Frankenheim}
\affiliation{%
Technische Universit\"at Darmstadt, Schlossgartenstrasse 8, D-64289 Darmstadt, Germany
}%

\date{\today}

\maketitle

This Supplemental Material to \textit{Threedimensional Alternating-Phase Focusing for Dielectric-Laser Accelerators} details (1) the determination of the individual structure constants and their dependencies on various parameters, (2) the shaping of the required laser pulses, (3) the accelerator design for 2.5 keV injection energy, and (4) a summary of the available video files.

\section{Structure constant and Bandwidth}
\begin{figure}[h]
\includegraphics[width=0.6\textwidth]{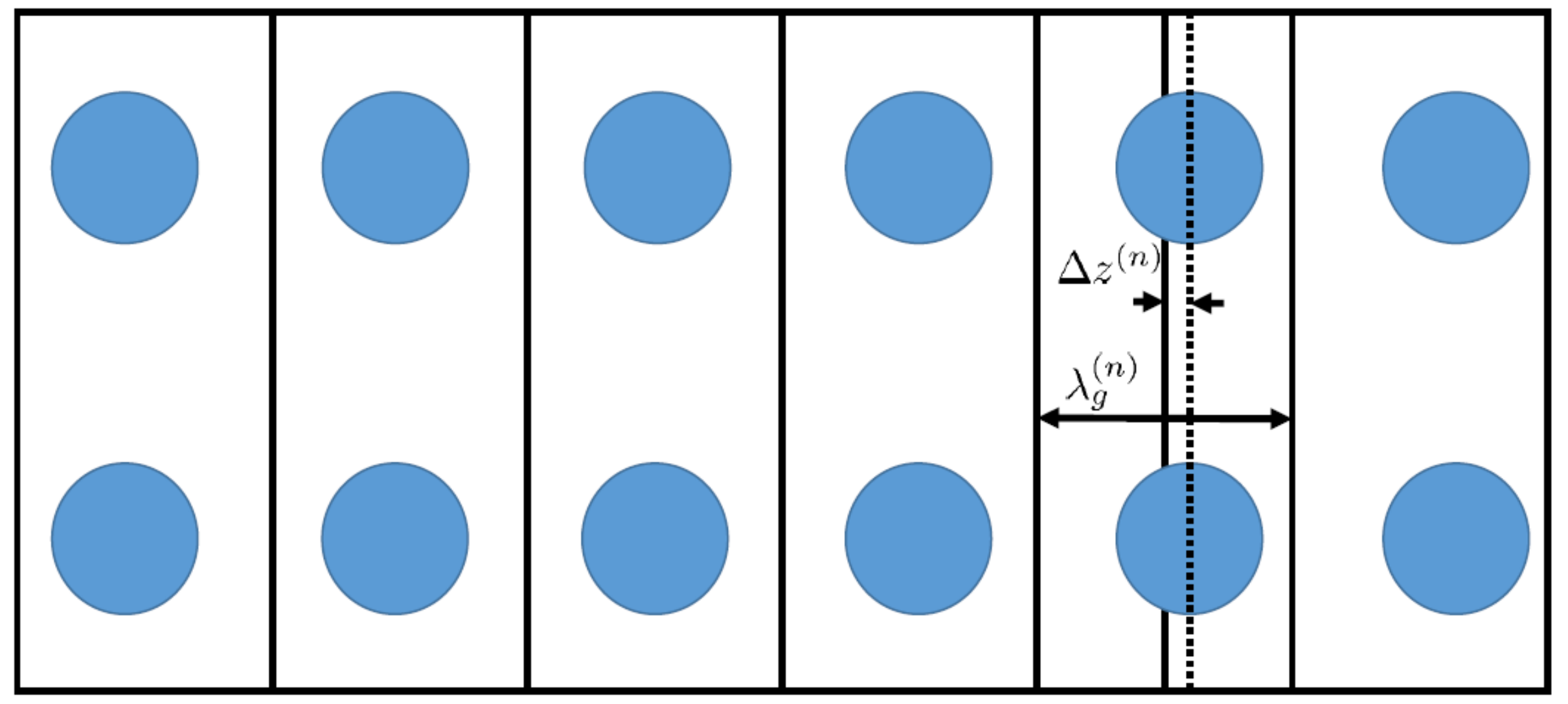}
\caption{Correcting the phase shift due to the cell length chirp allows to keep the pillar shape identical.}
\label{Fig:shift}
\end{figure}

In the literature, the acceleration on axis peak gradient is often referred to as a structure constant multiplied by the incident laser field strength $E_L$. We denote that by $|e_{10}|$, where the first mode coefficient is defined as~\cite{Niedermayer2017BeamScheme}
\begin{equation}
        e_{10}=\frac{1}{\lambda_g}\int_{-\lambda_g/2}^{\lambda_g/2}
    E_z(0,0,z)e^{-i \frac{2\pi}{\lambda_g} z} \dz
\end{equation}
with $E_z$ being the longitudinal component of the electric field in the frequency domain, i.e. at the fixed frequency $f_0=c/\lambda$.
We adjust the transverse coordinate system such that $\nabla_\perp e_{10}(0,0)=0$, i.e., the origin is at the saddle point for the counter-phase scheme and at the minimum for the in-phase scheme.
The structure constant is then $|e_{10}^\mathrm{norm}|=|e_{10}/E_L|$. Knowledge of the phase of $e_{10}$ is required to precisely inject the electron beam at proper time, i.e. to design an attosecond buncher~\cite{Niedermayer2017DesigningChip,Niedermayer2018Alternating-PhaseAcceleration,Black2019NetAccelerator,Schonenberger2019GenerationAcceleration} that can inject the electron beam at the synchronous phase. Phase adjustments in $\arg(e_{10})$ can be performed by moving the pillar away from the DLA cell center by $\Delta z^{(n)}=\lambda_g \Delta\arg(e_{10}) /2\pi$, see Fig.~\ref{Fig:shift}. This allows in particular to compensate for phase drifts when using the identical pillar shapes for a continuum of values of $\lambda_g^{(n)}=\lambda\beta^{(n)}$. 

In order to allow deviations from a strictly $\lambda_g$-periodic grating, in particular for the APF phase jumps, each grating cell has to support a certain bandwidth $\Delta f$. This means we have to optimize the structures for a \textit{low} quality factor $Q$ and a high $|e_{10}|$, similar to the geometry optimizations in conventional RF accelerator cavities, where first the geometry is optimized is to maximize $R/Q$ ($R$ is the shunt impedance) and later the surface is optimized to obtain highest $Q$. In this way, maximal $R=(R/Q)Q$ is obtained. For DLAs, only the geometry is subject to optimization and $|e_{10}|/Q$ is pragmatically optimized by first maximizing $|e_{10}|$ and then simply checking if sufficient bandwidth is supported, i.e. if $Q$ is sufficiently small.

\begin{figure}[bh]
\includegraphics[width=1.0\textwidth]{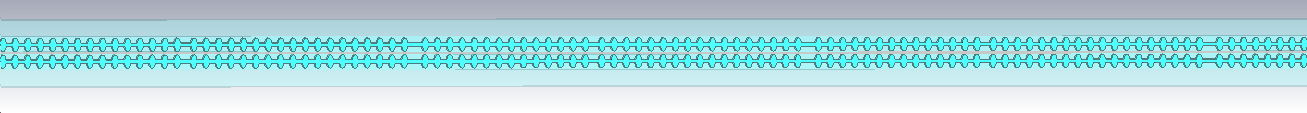}
\includegraphics[width=1.0\textwidth]{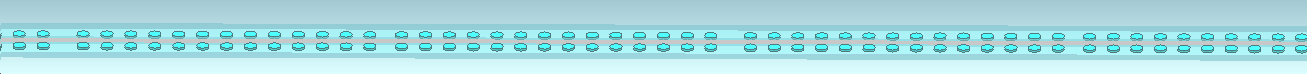}
\includegraphics[width=0.85\textwidth]{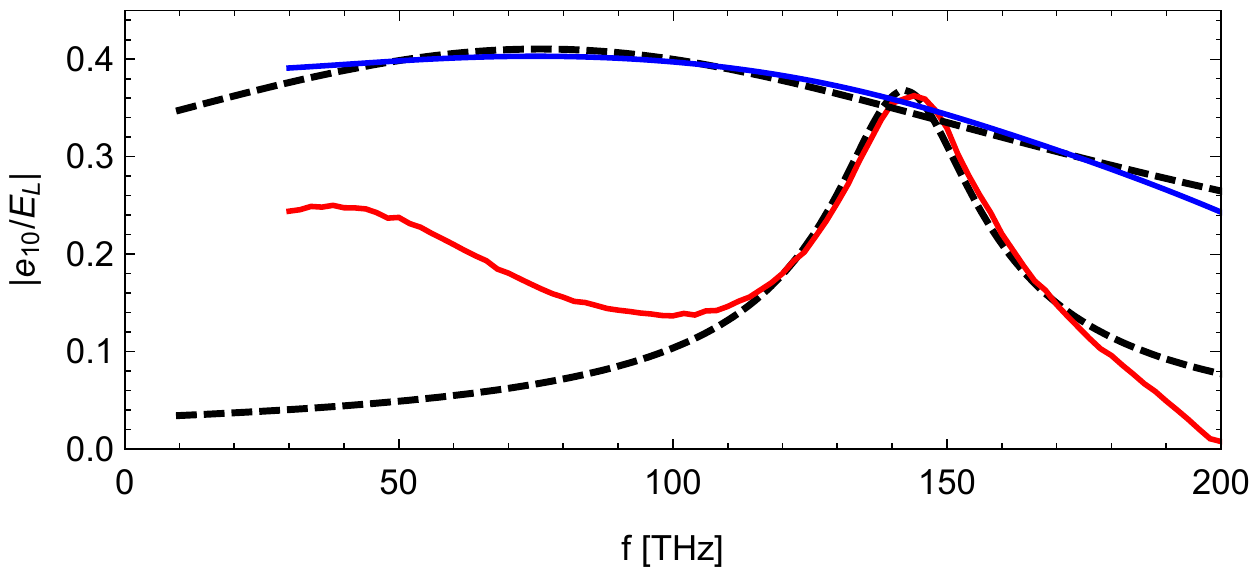}
\caption{Comparison of the bandwidth $\Delta f= f_0/Q$ of a connected (upper drawing, red curve) and not connected structure (lower drawing, blue curve). For the connected structure the quality factor is $Q=11.5$, resulting in 12.4~THz bandwidth. For the individual pillars, we find $Q=0.7$ and 104.8~THz bandwidth. The displaced resonance frequency ($f_r\approx75$~THz) is not an issue due to the large bandwidth.}
\label{Fig:BW}
\end{figure}

We determine the bandwidth of a particular structure by fitting $e_{10}(f)$ to a Lorentzian peak, while the Wideroe condition $\lambda_g=\beta\lambda$ is always fulfilled. Thus, while sweeping the frequency, the particle velocity has to be adjusted according to $\beta=\lambda_g f/c$ and the proper frequency dependent definition of $e_{10}$ reads
\begin{equation}
    e_{10}(f)=\frac{1}{\lambda_g}\int_{-\lambda_g/2}^{\lambda_g/2}
    E_z(0,0,z;f)e^{-i \frac{2\pi}{\lambda_g} z} \dz.
    \label{Eq:structF}
\end{equation}
The resonance frequency $f_r$, the peak amplitude $|e_{10}(f_r)|$ and the quality factor $Q$ are determined by fitting the frequency sweep to the Lorentzian
\begin{equation}
    |e_{10}(f)|=\frac{|e_{10}(f_r)|}{\sqrt{1+Q^2(\frac{f-f_r}{f_r})^2}}.
\end{equation}
The resulting resonance curves and quality factors for a are plotted in Fig.~\ref{Fig:BW}. As examples, we compare a structure made of individual pillars and a structure which is topologically connected. Both structures have been geometrically optimized to maximize $e_{10}$ at $f_0=150$~THz.
Note that some structures might exhibit multiple resonances, that have to be fitted individually to Lorentzian peaks, however, here we look only on those close to the excitation center laser frequency $f_0$.
Moreover, especially at high bandwidth, $f_r$ and $f_0$ might differ significantly.

We observe, that the connected structure does not provide sufficient bandwidth to support APF phase jumps.
A smooth curve for $|e_{10}|$, as in Fig.~6 of the main manuscript, is not attained for the connected structure, since the phase jumps induce wild oscillations. Thus we reject the connected structures and keep the individual pillars. As a coarse criterion for a single cell that supports APF phase jumps we formulate $Q\lesssim 1$.

The individual pillars however cannot be bulk grounded and require metal atomic layer deposition (ALD) to slowly remove charge that accumulated due to electron beam loss.
The calculation of $e_{10}(f)$ can be performed either as a frequency sweep and point-wise evaluation of Eq.~\ref{Eq:structF} or at once as a time domain broadband pulse excitation with simultaneous evaluation of multiple frequencies.

\section{Laser Pulse Shape}
A fully scalable DLA requires a tilted laser pulse, such that the time of interaction of the laser pulse with the electron beam becomes independent of the time of interaction with a particular pillar of the structure. The tilt angle is given by (see e.g.~\cite{Wei2017Dual-gratingLaser})
\begin{equation}
    \alpha=\arctan\frac{1}{\beta}
\end{equation}
and can be implemented e.g. by a diffraction grating.
The laser field amplitude follows a bi-Gaussian distribution as
\begin{equation}
E_z\propto \exp -\frac12{\left[ 
\left( \frac{(z-z^*)\cos\alpha+(y-\tilde{y})\sin\alpha}{\sigma_z} \right)^2 
+\left( \frac{(z-z^*)\sin\alpha-(y-\tilde{y})\cos\alpha}{\sigma_y} \right)^2
\right] },
    \label{Eq:Amplitude}
\end{equation}
where $\sigma_y=c\sigma_t$ is the pulse length and $\sigma_z$ is the pulse width, which has to cover the entire DLA structure.
The polarization is in $z$-direction and the phase fronts are flat $xz$-planes.
For constant reference velocity $\beta$, the laser pulse can be arbitrary short.
However, when the electron is accelerated, its trajectory within the laser pulse will not be linear anymore. Ideally, the tilted pulse would be replaced by a  "banana"-shaped pulse, in order to exactly follow the acceleration ramp. A constantly tilted pulse can however approximate the "banana" if the interaction is over a finite length $L$. The drawback of this is that a minimum length of the pulse is required.

In order to calculate the minimal pulse length and the optimal tilt angle, we have to compute the trajectory $y(z)$ of the electron within the laser pulse. Its derivative provides  
\begin{equation}
    \frac{\dy}{\dz}=\tan \alpha(z) = \frac{1}{\beta(z)} \;\;\;\Rightarrow\;\;\;
    y(z)= \int_0^z \frac{\d \tilde z}{\beta(\tilde z)}.
    \label{Eq:Inegral}
\end{equation}
A secant to this trajectory is 
\begin{equation}
    y_s(z)=\frac{y(L)}{L}z
\end{equation}
where $L$ is the length of the accelerator. 
The optimal tilt angle $\alpha^*=\arctan\frac{1}{\beta(z^*)}$ of a linearly tilted pulse is now found at the position where the difference between the "banana" and its secant is maximal, i.e.
\begin{equation}
    z^*=\mathrm{argmax}\;\;  y(z)-y_s(z).
\end{equation}
An estimate of the required pulse length, such that at perfect timing the laser amplitude has dropped not more than to $\exp(-\xi^2/2)$ of the peak value, is 
\begin{equation}
    \sigma_y=\frac{\cos(\alpha^*)}{\xi} \max y(z)-y_s(z) 
\end{equation}
under the condition of $\sigma_z \gg \sigma_y$.
The proper value for $\tilde{y}$ is found as the middle between $y(z^*)$ and $y_s(z^*)$ as $\tilde{y}=\frac12 \left[y(z^*)+y_s(z^*)\right]$.

In the following we take $\xi=1$ and a linear energy ramp, i.e., $W(z)=\gamma_0 m_ec^2+Gz$, with $\gamma_0$ being the initial Lorentz factor and $G$ being the constant gradient. The integral~\ref{Eq:Inegral} is solved as
\begin{equation}
    y(z)=\frac{m_ec^2}{G}\left[\sqrt{\left(\gamma_0+\frac{Gz}{m_ec^2}\right)^2-1}               -\sqrt{\gamma_0^2-1}\right]
    =\frac{m_ec^2}{G} \left[\beta(z)\gamma(z)  -\beta_0\gamma_0\right] 
\label{Eq:Banana}
\end{equation}
and $\beta(z^*)=L/y(L)$.
The optimal center point $z^*$ is obtained by 
setting $W(z^*)=m_ec^2 \gamma^*$ as
 \begin{equation}
     z^*=\frac{m_ec^2}{G}(\gamma^*-\gamma_0).
 \end{equation}
The pulse length is thus explicitly 
\begin{equation}
    \sigma_y=\frac{m_e c^2}{2G}\left[
    \beta^*\gamma^*-\beta_0\gamma_0-(\gamma^*-\gamma_0)/\beta^*  \right].
\end{equation}

\begin{figure}[h]
\includegraphics[width=0.6\textwidth]{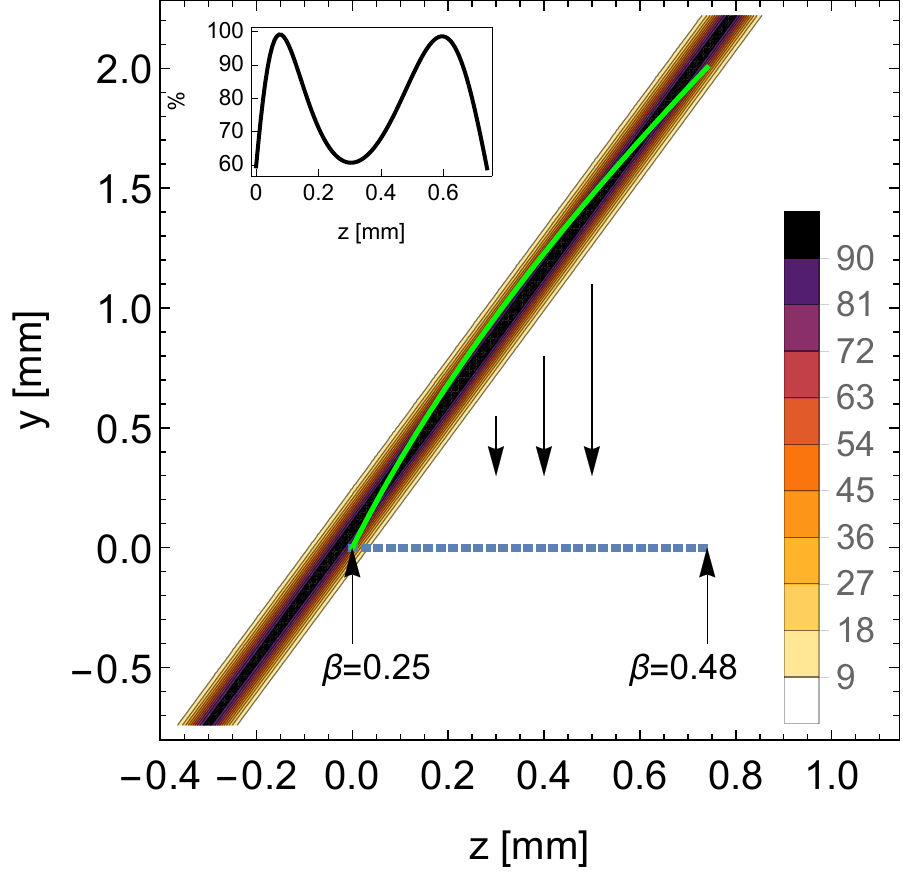}
\caption{Contours of $E_z$ as Eq.~\ref{Eq:Amplitude} for the values $\alpha^*=69.72^\circ$, $\sigma_t=179$~fs, $\sigma_z=1$~cm in percent. The dashed blue line represents the accelerator structure. The green line shows the trajectory of the electron within the pulse according to Eq.~\ref{Eq:Banana}. The inset is a parametric evaluation of the laser amplitude on the electron trajectory.}
\label{Fig:Tilt}
\end{figure}

The pulse parameters for the accelerator in the main paper, i.e. for $\beta=0.25$ to $\beta=0.48$ in 740\,$\mu$m, i.e. a 73~MeV/m average gradient, are depicted in Fig~\ref{Fig:Tilt}. The optimal tilt angle is $69.72^\circ$ and a pulse duration of $\sigma_t=179$~fs is required.
The damage threshold for Silicon is about 2~GV/m for 100~fs pulses at $\lambda=1.06\,\mu$m~\cite{Pronko1998AvalanchePulses}. Soong et al.~\cite{Soong2012LaserAccelerators} indicate that
the damage threshold does not vary more than a factor
of 2 between $1\,\mu$m and $2\,\mu$m, therefore we assume to be close but below the damage threshold when the amplitude is 500~MV/m from both lateral sides. We furthermore assume that SOI structures have the damage threshold of silicon, since the one of the oxide is significantly higher.
 
This damage threshold fluence constraint requires pulses as short as 100 fs, which in turn requires creating the "banana"-shape as described by Eq.~\ref{Eq:Banana}. A possible technique for this is given by the combination of a Spatial Light Modulator (SLM) and a Deformable Mirror (DM), see~\cite{Sun2015PulsePulses}. Another option would be to split the pulse in parts with individual tilt angles, such that the "banana" is approximated by linear pieces.

The 40\% lower amplitudes at the electron trajectory can be compensated by increasing the overall laser power within the damage constraint. In contrast to lower than nominal laser amplitudes, slightly higher amplitudes do not lead to electron beam loss, since increased $K$.values lead to decreased $\hat\beta$ functions. On the other hand, however, the $\hat\gamma$ function is increased, leading to a larger energy spread (and also angle spread) of the outcoming electron beam. The reference output energy is not affected by slight laser amplitude changes, since it is hard-coded into the accelerator lattice design by the periodicity chirp.

\section{Ultra low injection energy design}
\begin{figure}[h]
    \centering
    \includegraphics[width=0.95\textwidth]{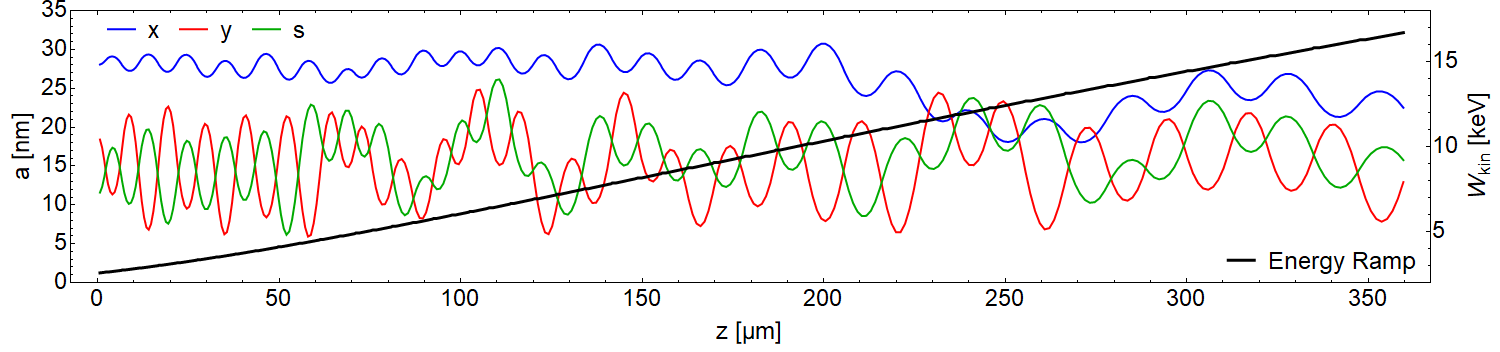}
        \includegraphics[width=\textwidth]{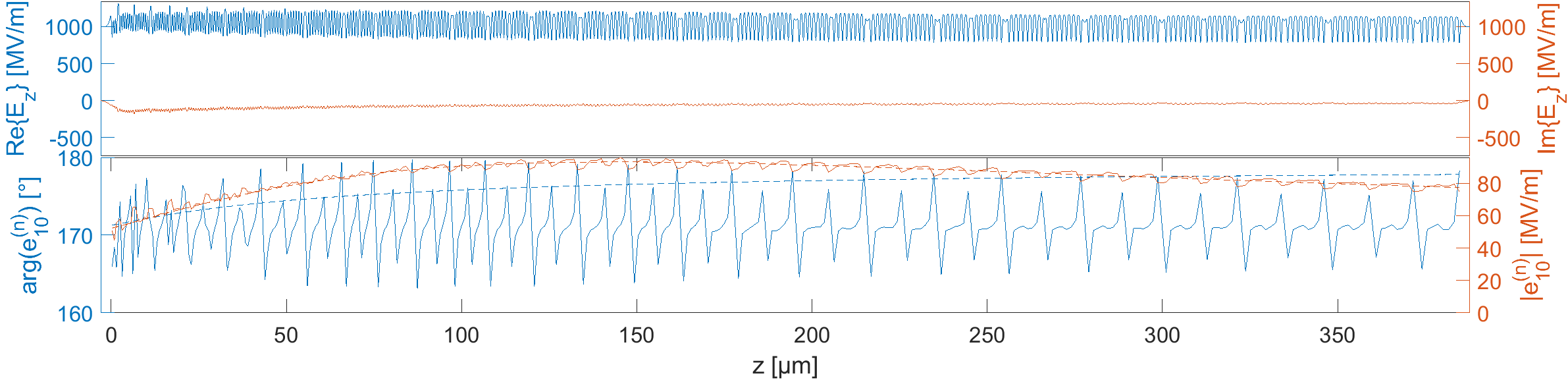}
    \caption{Upper panel (design): envelopes for $\eps_n=2.5\,$pm and kinetic energy ramp.
Lower Panel (3D analysis): complex electric field $E_z(0,0,z)$ and spatial Fourier coefficients $e_{10}^{(n)}$ calculated from the full field windowed in each DLA cell $n$. The dashed lines represent the individually computed values of $e_{10}$ under periodic boundary conditions, which were used for the design.}
    \label{Fig:LAtticeultralowE}
\end{figure}

Similarly to the design in the main paper, we show another accelerator design here at ultralow injection energy of 2.574~keV ($\beta=0.1$).
We choose a wavelength of $\lambda=6~\mu$m, at which silicon is still transparent, and the first period $\lambda_g^{(1)}=600$~nm is not too small for fabrication. 
The aperture is chosen as 500~nm and the dimensions of the identical pillars are $r_y=100$~nm and $r_z=250$~nm. The silicon layer height is 440~nm and the laser amplitude is again 500~MV/m from both lateral sides. The design as well as the analysis of the 3D field results are shown in Fig.~\ref{Fig:LAtticeultralowE}.

\begin{figure}[b]
    \centering
    \includegraphics[width=0.45\textwidth]{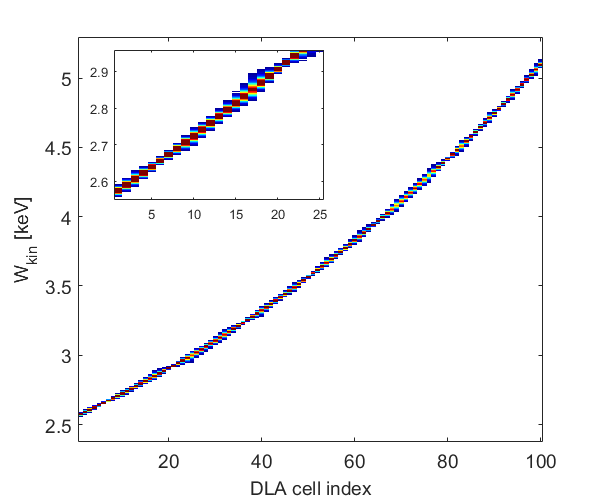}
    \includegraphics[width=0.45\textwidth]{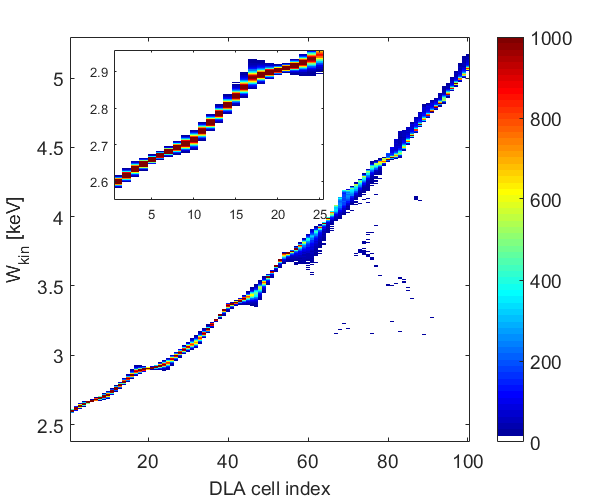}
    \caption{Spectrograms for the first 100 cells of the $\beta=0.1$ to $\beta=0.25$ accelerator from DLAtrack6D simulation. The left panel is at design injection energy and produces 94\% total throughput, the right panel is at +25~eV injection energy offset and produces only 35\% total throughput. The insets are enlargements of the first 25 cells.}
    \label{Spectrogram}
\end{figure}

Due to decreased bandwidth at low $\beta$, the deviation of the full 3D field amplitude and phase from the design values (periodic boundaries) is quite substantial.
Thus, further structure optimization is required in the future. Moreover, at such low injection energy, energy and phase mismatch have a large impact on the total particle transmission.

The particle tracking simulation was performed both with DLAtrack6D~\cite{Niedermayer2017BeamScheme} using $e_{10}^{(n)}$ and with CST~\cite{CST2019CSTSuite} taking into account the full fields. The emittance was taken as $\eps_x=10$~pm, $\eps_y=20$~pm and the bunch length was $\sigma_z=5$~nm with matched energy spread. The expected throughput should be in the order of 94$\%$, see Fig.~\ref{Spectrogram}, however with CST we obtain only about 20\%. We account this to two issues. First, the insufficient field quality especially at the lower end of the accelerator. Second, and likely more decisive, the CST simulation in full 3D frequency domain fields creates an artificial energy offset at injection.

When the particles are released from a window within the computational domain, they see the already present fields immediately. This violates the conditions of the Lawson-Woodward theorem, which states that a plane wave cannot produce first order net acceleration~\cite{England2014DielectricAccelerators}. However even a plane wave ($e_0$ mode) can do a net transfer of energy to the particles if they suddenly appear out of a window. This ends up in a significant injection energy error, in the order of 100~eV. In order to study this effect, we have artificially included it in a DLAtrack6D simulation. The effect of a 25~eV (1\%) offset is shown in Fig.~\ref{Spectrogram}, right panel. A coherent oscillation arises, which strongly probes the nonlinearities of the fields. This leads to emittance growth in all 3 planes and eventually beam loss, i.e. we obtain only 35\% througput in the example. Manual correction (fine-tuning) of the injection energy is possible up to an estimated residual of 10 to 20~eV. We note, that this is only an issue of the simulation, not the design.

In general, a lower limit for the injection energy is found by practical considerations. These are in  particular related to fitting the envelope $a$ into the  aperture $A$ of the structure. The scaling laws for the envelope are 
\begin{equation}
a= \sqrt{\hat\beta \eps}\propto \sqrt{\frac{1}{K} \frac{\eps_n}{\beta\gamma}}\propto\frac{\beta\gamma}{\sqrt{|e_{10}|}},
\end{equation}
i.e. due to the strong $K$ values at low energy, an at most quadratic drop of the structure constant with $\beta$ is tolerable (without changing the aperture).

\section{Online available video examples}
Three videos are online available with this supplemental material to be downloaded. They represent DLAtrack6D simulations for the case in the main paper, the ultra low energy example in this supplement, and the +25~keV injection energy offset example of the latter.
Scatter plots are intended to draw the reader's attention to the high amplitude particles, which perform threedimensionally coupled nonlinear oscillations.

\bibliography{./references}